# Controlling light in complex media beyond the acoustic diffraction-limit using the acousto-optic transmission matrix

Ori Katz[1,2,3,*], François Ramaz[2], Sylvain Gigan[3], Mathias Fink[2,*]

[1] Department of Applied Physics, The Selim and Rachel Benin School of Computer Science & Engineering, The Hebrew University of Jerusalem, Jerusalem 9190401, Israel

[2] Institut Langevin, ESPCI Paris, PSL Research University, CNRS UMR7587, INSERM U979, UPMC, 1 rue Jussieu, 75005 Paris, France.

[3] Laboratoire Kastler Brossel, Université Pierre et Marie Curie, Ecole Normale Supérieure, Collège de France, CNRS UMR 8552, 24 rue Lhomond,75005 Paris, France

[*] e-mail: orik@mail.huji.ac.il, mathias.fink@espci.fr

**Studying the internal structure of complex samples with light is an important task, but a difficult challenge due to light scattering. While the complex optical distortions induced by multiple scattering can be effectively undone with the knowledge of the medium's scattering-matrix, this matrix is generally unknown, and cannot be measured with high resolution without the presence of fluorescent or absorbing probes at all points of interest. To overcome these limitations, we introduce here the concept of the acousto-optic transmission matrix (AOTM). Taking advantage of the near scattering-free propagation of ultrasound in complex samples, we noninvasively measure an ultrasonically-encoded, spatially-resolved, optical scattering-matrix. We demonstrate that a singular value decomposition analysis of the AOTM, acquired using a single or multiple ultrasonic beams, allows controlled optical focusing beyond the acoustic diffraction limit in scattering media. Our approach provides a generalized framework for analyzing acousto-optical experiments, and for noninvasive, high-resolution study of complex media.**

Conventional focusing and imaging techniques based on the Born approximation generally fail in strongly scattering media because of the *multiple* scattering events that any incident optical beam undergoes. However, multiple light scattering does not lead to an irretrievable loss of information.. The complex wavefront distortions, even deep inside diffusive samples, can be effectively reversed by high-resolution shaping of the input optical wavefront[1], in a fashion analogous to time-reversal experiments in ultrasound[2]. The ability to digitally control optical interference in multiply scattering media has recently given rise to new focusing and imaging techniques[3, 4]. Following the pioneering work of Vellekoop and Mosk[5], spatial light modulators (SLM) were used for correcting spatial[6-9], temporal[10, 11], spectral[12] and polarization[13] distortions, and to optimize transmission[14] through multiply scattering media, via wavefront optimization. A generalized theoretical framework underlying all experiments involving light propagation in complex media is the scattering/transmission matrix (TM) formalism[6, 15]. The TM essentially contains the medium's response at every output spatial mode to excitation at any input spatial mode (the medium's Green function). The experimental access to the TM was first made possible by measuring the medium's response from each pixel of an SLM to each pixel of a camera placed at the desired output plane[6]. However, while the experimentally measured TM allowed taking advantage of multiple scattering for focusing and imaging, this could only be done at the camera detection plane, severely limiting the practical interest in biomedical applications, where the goal is to image *inside* a complex sample, without direct access to the target. To control scattered light inside scattering samples noninvasively, a 'guide star' providing feedback on the optical intensity at the target point is required[4]. Recent exciting developments along this path include fluorescence markers[16], non-linear optical particles[17-19], photoacoustic feedback[20-22], or acousto-optic tagging[23-29]. The last two techniques are multi-wave approaches[4], which benefit from the interaction between light and sound, potentially combining the best of the two waves: the near scattering-free propagation of ultrasound and the selective contrast, and sub-micron diffraction-limited resolution of optical waves. In photo-acoustics, the interaction of light with absorbing parts of the sample generates ultrasound, which allows ultrasonic imaging of optical absorbers, or ultrasound-guided focusing on these absorbers. In acousto-optic imaging, the requirement for optical absorption is removed by exploiting frequency modulation of the diffused light by an ultrasound beam that is tightly focused inside the sample[30]. Detection of the frequency-shifted light enables observing only the ultrasonically 'tagged' portions of the optical wave, which have travelled through the ultrasonic focal spot. Scanning the ultrasonic focus can thus provide a mapping of the optical intensity distribution inside the sample, with the spatial resolution of the ultrasonic focus. The ultrasonically tagged light can also be focused back to the ultrasonic focus via phase-conjugation, in time-reversed ultrasonically encoded (TRUE) optical focusing[23, 24, 26], or via iterative optimization[29]. The main drawback of acousto-optic based approaches is that the spatial dimensions of the ultrasound focus are of the order of the ultrasonic wavelength. This yields an imaging resolution that is orders of magnitude lower than the optical diffraction limit, and an optical focus containing a large number of optical speckles with a relatively low peak to background intensity ratio[4].

Overcoming the acoustic diffraction limit in acousto-optics is of major interest for many applications. To date two approaches to overcome the acoustic diffraction-limit in acousto-optics have been proposed: iterative TRUE (iTRUE)[27, 28], and time reversal of variance-encoded (TROVE) optical focusing[25]. In TROVE one analyzes fluctuations of ultrasonically tagged light intensity for different random inputs, and computes an optical wavefront that focuses to the location with increased intensity fluctuations variance, allowing in principle optical-speckle size focusing[25]. In iTRUE, multiple iterations of phase conjugation operation are used to improve the focusing resolution[27, 28]. At each iteration the optical beam refocuses to the ultrasonically tagged region, spatially encoded again by the ultrasound focal spot pressure distribution, and shrinks progressively. For an ultrasonic spot having a Gaussian profile, performing $N$ iteration of phase conjugation provides a $\sqrt{N}$ resolution increase beyond the acoustic diffraction limit. Both iTRUE and TROVE rely on digital optical phase-conjugation (DOPC), which requires a very precise pixel-to-pixel alignment of a high-resolution SLM and camera, which can be experimentally challenging to maintain[31].

Here, we introduce a novel generalized concept for optical measurement and control using ultrasound tagging: the acousto-optic transmission matrix (AOTM). Our concept is based on the understanding that any measurement that utilizes linear ultrasound-tagging can be described by a single linear operator, and that this operator can be described by a single matrix. Thus, we show that a single AOTM provides a general, concise, and full description of light propagation in any acousto-optic experiment. We experimentally demonstrate how the AOTM can be measured using a single or multiple ultrasonic focused beams, and how it can be computationally processed to provide sub-acoustic optical focusing inside a complex medium, without requiring a DOPC system, i.e. using any positioning of an SLM and a camera. Since the AOTM describes any acousto-optic experiment, we show how TRUE, iTRUE, and TROVE can be described and compared under the AOTM framework, and how the AOTM allows to overcome some of the limitations of these approaches.

**Principle of the AOTM with a single ultrasonic beam**

Consider a general acousto-optical experiment, such as the one schematically depicted in Figure 1a, where diffused quasi-monochromatic light is ultrasonically tagged by a focused ultrasound beam, and subsequently measured by a camera placed outside the medium. Assuming linear light propagation and linear acousto-optic interaction, the relationship between any input optical field $E^{in}(f_o)$ at the optical frequency $f_o$, and the ultrasonically-tagged output field at the camera plane, $E^{out}(f_o + f_{US})$, which is frequency shifted to a frequency $f_o+f_{US}$ by the ultrasound beam at frequency $f_{US}$, is given by a linear operator $T$:

$$E^{out}(f_o + f_{US}) = TE^{in}(f_o) \qquad (1)$$

We define the matrix describing this operator in the spatial domain as the AOTM. Specifically, each element of the AOTM, $t_{mn}$, gives the complex amplitude of the acoustically-tagged optical field at the output spatial position $r_m$, $E_m^{out}$, as result of an input field at position $r_n$, $E_n^{in}$ (an acousto-optic 'Green function'):

$$E_m^{out}(f_o + f_{US})|_{input\ field\ at\ r_n} = t_{mn}E_n^{in}(f_o) \quad (2)$$

As result of linearity, the output acoustically-tagged field measured at $r_m$ for a general input field is given by:

$$E_m^{out}(f_o + f_{US}) = \sum_n t_{mn}E_n^{in}(f_o) \quad (3)$$

where the summation is over all spatial input modes, $n$. Figure 1a depicts the basic setup required for experimentally measuring the AOTM and consequently exploiting it for controlled focusing inside a complex medium. The setup is based on the well-established approach for measuring the optical transmission-matrix (TM)[6], with the addition of an ultrasound transducer that generates an ultrasound pulsed focus inside the medium, and a reference arm for off-axis holography. The setup is composed of an illuminating laser beam that passes through a computer-controlled SLM to provide injection of controlled optical modes into the medium. The acoustically-tagged output scattered field at frequency $f_o + f_{US}$ is measured outside the medium via off-axis, phase-shifting holography[32], using a pulsed plane-wave reference beam that is synchronized with the ultrasound pulses (see Methods).

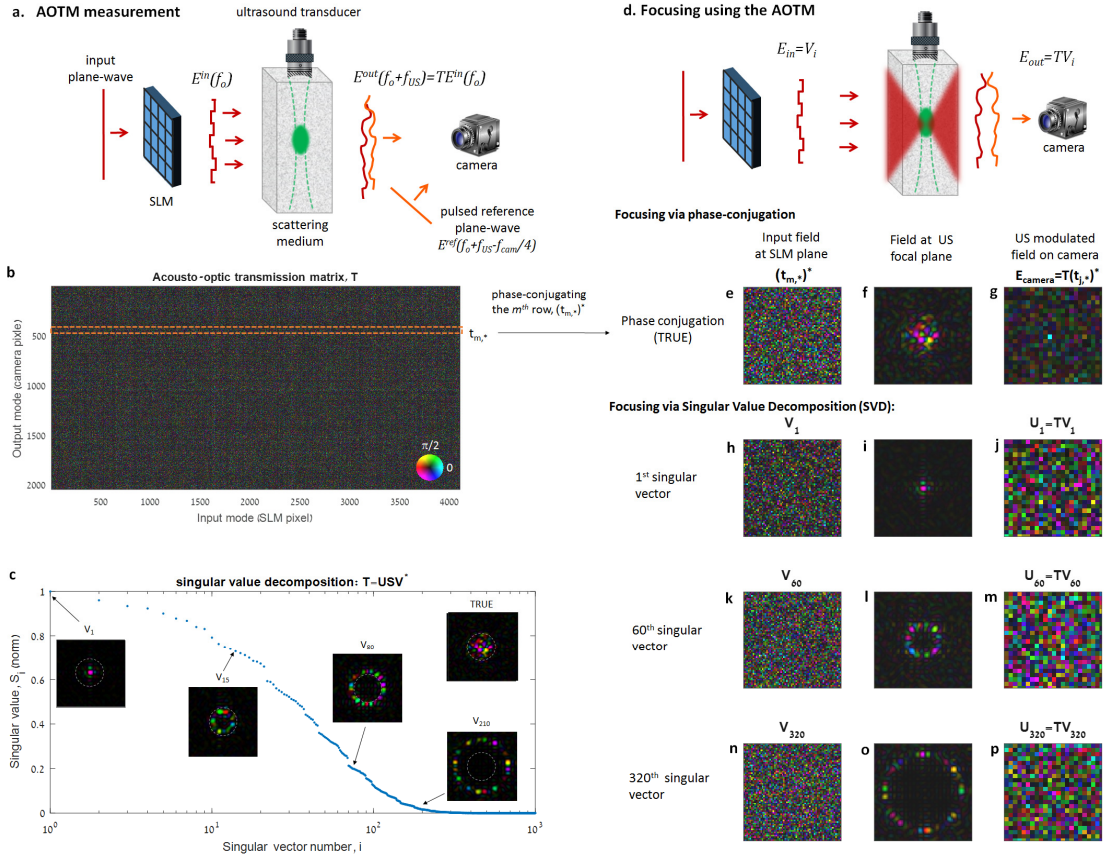

**Figure 1: Acousto-optic transmission matrix (AOTM) measurement and use for sub-acoustic light control** (numerical results). **a,** Setup for measuring the AOTM. **b**, A measured AOTM, **c,** Singular values of the AOTM (insets: optical fields at the ultrasound focal plane when the corresponding singular vectors are injected into the medium), **(d-p),** using the AOTM for light control: **e-g,** phase-conjugating the $m^{th}$ row of the AOTM is equivalent to TRUE focusing, enhancing the optical intensity inside the ultrasound focus; **h-j,** injecting the AOTM first singular vector focuses light at the center of the ultrasound focus, with a resolution beyond the acoustic diffraction limit, equivalent to infinite iterations of iterative TRUE; **k-p,** injecting singular vectors with lower singular values result in intensity enhanced rings with increasing diameter around the ultrasound focus. Low singular values result in increase in intensity outside the ultrasound focus, and reduced optical intensity inside it (n-p).

The AOTM is measured by sequentially injecting each of the $n=1..N_{SLM}$ input modes, $E_n^{in}(f_o)$, at the laser frequency $f_o$ into the medium (where $N_{SLM}$ is the number of SLM pixels used), and measuring the ultrasonically tagged, frequency-shifted, output field $E_m^{out}(f_o + f_{US})$ at each of the $m=1..M$ camera pixels simultaneously. Following Equation (3), in each of the $n=1..N_{SLM}$ measurements steps, the $n^{th}$ column of $T$ is acquired. The AOTM measured with this setup thus describes the propagation of light from the input plane of the SLM, *through the small ultrasound focus*, to the camera plane. The size of $T$ is $M \times N_{SLM}$, when a single ultrasound focus is used (Fig.1b). While a conventional TM, measured without the ultrasound focus, reflects the combined interference of all optical paths inside the scattering volume, the spatially localized ultrasound focus allows spatially resolved probing and light control in a specific volume inside the medium using the AOTM, as we show below.

Once the AOTM has been measured (Fig.1b), it can be used for optical focusing inside the medium by several different approaches (Fig.1d-p). The most straightforward but least powerful approach is via direct phase-conjugation: displaying the phase conjugate of the $m^{th}$ row of the AOTM on the SLM (Fig.1b inset, Fig.1e), leads to light focusing at the $m^{th}$ camera pixel (Fig.1g). However, as result of the ultrasound tagging, this externally focused wave also concentrates optical intensity inside the acoustic focus (Fig.1f). The dimensions of the intensity-enhanced volume inside the sample are given by the ultrasound focus, and are dictated by acoustic diffraction. This is the same limit as in TRUE-focusing experiments[23, 24, 26], which are based as well on phase-conjugation. This is not a coincidence, but rather a deeper result obtained from optical reciprocity: the mathematical description of any TRUE experiment is a phase-conjugation of an AOTM row corresponding to initial illumination from the $m^{th}$ camera pixel (see Supplementary Section 2).

A considerably more powerful optical focusing and control approach that is enabled by the AOTM and allows surpassing the acoustic diffraction limit is via singular value decomposition (SVD) of the AOTM. SVD is a powerful tool in matrix analysis, and was recently used to identify transmission-eigenchannels[33-36], and for selective focusing[21, 37]. The main interest in the SVD of the AOTM is that, by definition, the singular vectors of the AOTM, are the eigenvectors of $T^H T$ (where $T^H$ denotes the Hermitian conjugate of $T$). $T^H T$ is the matrix describing the operator of consecutive two iterations of an iterative phase conjugation (iTRUE) experiment (see Supplementary section 3). An intuitive explanation for this fact can be obtained by considering that $TE^{in}$ describes a single pass of an input field $E^{in}$ through the medium and ultrasound focus, and thus after a first phase-conjugation step and back-propagation through the medium the measured field in the second iTRUE iteration would be $T^T(TE^{in})^*$, which after consecutive phase conjugation would be $(T^H T)E^{in}$. Thus, the input wavefront at the $2k^{th}$ iteration of iTRUE is $(T^H T)^k E^{in}$. The desired wavefront required for the tightest optical focus is obtained for $N \rightarrow \infty$, and is given by the eigenvector of $T^H T$ having the largest eigenvalue. This eigenvector is, by definition, the first singular vector of the AOTM, $T$. Using the SLM to inject this mode into the medium results in the most tightly focused optical spot at the center of the acoustic focus (Fig.1h-j). The optical focus size reaches the optical diffraction limit (a single optical speckle grain) if a sufficiently large number of input matrix modes, $N_{SLM}$, are measured. This number is larger, the larger is the number of optical modes (speckles) contained in the acoustic focus (see Supplementary section 5). Not only the first singular vector of the AOTM is of interest: singular vectors with decreasing singular values will form concentric rings with increasing diameter around the center of the ultrasound focus (Fig.1k-p). Using even lower singular values leads to concentration of energy *outside* the acoustic focus, in a fashion resembling open channels in systems containing localized absorption[36]. This resemblance may be understood from the fact that the singular values of the AOTM represent the energy transmission through the acoustic focus, while singular values of the TM represent energy transmission of transmission eigenchannels through the complex medium. The distribution of singular values of the AOTM (Fig.1c) is affected by the shape of the acoustic focus, and the number of optical modes contained within the acoustic focus. For the Gaussian ultrasound focus considered here, the singular values gradually decrease, while for a theoretical 'top-hat' circular tagging area they are abruptly cut at the number of optical modes contained in the acoustic focus (see Discussion and Supplementary section 4).

## Results

### AOTM using a single ultrasonic beam: Experiments

To experimentally demonstrate our approach we used the setup schematically described in Figure 2a, and described in detail in Supplementary Section 1 (see Methods). In this proof of principle, we used two thin diffusers as the complex sample, such that the optical fields inside the ultrasound focus could be directly inspected by removing the second diffuser after the AOTM measurement. We note that the approach is general and is not limited to thin scattering layers, as is proved numerically in the numerical results of Figure 1, where a multiply-scattering medium was considered (see Methods). A pulsed focused ultrasound transducer with a central frequency of 15MHz placed perpendicular to the light propagation direction was used for acousto-optic tagging. Figure 2 presents the results of optical control using the SVD of an experimentally measured AOTM. As expected, injecting the first singular vectors generates a sharp optical focus with dimensions smaller than the acoustic focus (Fig.2c-d). Injecting singular vectors with lower singular values results in the formation of concentric rings of increased optical intensity around the center of the ultrasound focus (Fig 2e-f). Using lower singular values leads to concentration of energy *outside* the acoustic focus (Fig.2g-i). The full-width at half-max (FWHM) transverse dimensions of the ultrasound focus (Fig.2b) and the sharpest focus formed via SVD of the AOTM (Fig.2c) are 170μm±10μm, and 35μm±5μm, respectively. The axial dimensions of these foci are 175μm±20μm, and 35μm±5μm, respectively. Thus, the SVD of the AOTM provides here a resolution increase of approximately 4.8.

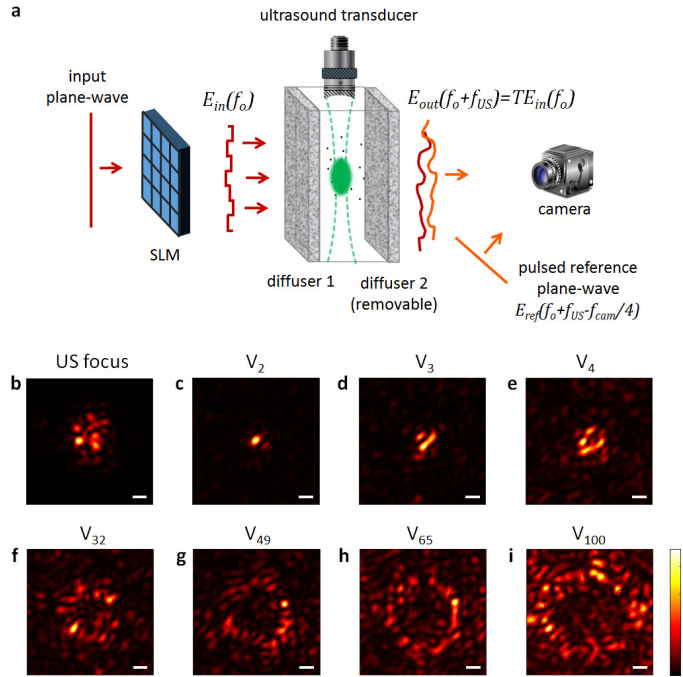

**Figure 2 Experimental results of SVD of the AOTM measured with a single ultrasound focus**. (a) Experimental setup. **(b-i),** measured optical intensity distributions at the ultrasound focal plane: **b,** optical intensity of the ultrasonically tagged field for a plane wave input, giving the size of the ultrasound focus. **c-i** optical intensity distributions obtained when injecting the 2nd, 3rd, 4th, 32nd, 49th, 65th, and 100th singular vectors of the AOTM. While the first singular vectors focus light at a sharp focus smaller than the ultrasound focus (c-d), singular vectors with lower singular values result in concentric rings around the ultrasound focus center with increasing diameter. Low singular values concentrate optical intensity outside the acoustic focus (g-i). scale-bars, 67μm.

## AOTM using multiple ultrasound beams

The position of the tight focus obtained using the SVD of the AOTM (Figs.1-2) is limited to the center of the ultrasonic focus because of its inherent symmetry. Here, we generalize the concept of the AOTM to a diversity of ultrasonic beams, and show that scanning a tight focal spot over multiple positions is possible by joint analysis of only two AOTMs measured with two ultrasonic beams.

As a simple example, we consider the separate measurement of two AOTMs, $T_1$ and $T_2$, using two different ultrasound focused beam, which are spatially shifted such that the two ultrasound focal spots, $P_1(x,y)$ and $P_2(x,y)$, partially overlap (Fig.3a-d). Such ultrasound focal spots are easily obtained by changing the time-delay $\Delta\tau=z/v_{US}$, between the ultrasound pulse and the pulsed optical illumination or reference wave (Fig.3a), where $z$ is the axial distance of the acoustic focal spot from the ultrasound transducer and $v_{US}$ is the speed of sound in the medium. Using the single AOTM SVD focusing approach of Figs.1-2, injecting the first singular vector of each of $T_1$ or $T_2$ separately, would form a tight optical focus only at the center of each of the ultrasound foci. However, the joint information in the two matrices can be exploited to scan the sharp focus along the axis connecting the centers of the two acoustic spots (Fig3i-k). One approach to obtain this is by injecting into the medium the first singular vectors of the matrix:

$$A_\alpha = \left[(T_1 - \alpha T_2)^H (T_1 - \alpha T_2)\right]^{-1} \left[(T_1 + \alpha T_2)^H (T_1 + \alpha T_2)\right] \quad (4)$$

where $\alpha$ is a positive weighting parameter controlling the focus position. $\alpha=1$ yields focusing at the middle of the line connecting the two ultrasound foci centers (Fig.3i). The matrix $A_\alpha$ is the multiplication between the TRO of the weighted sum of $T_1$ and $T_2$: $T_{1+\alpha 2} = (T_1 + \alpha T_2)$, by the inverse of their weighted difference: $T_{1-\alpha 2} = (T_1 - \alpha T_2)$. Scanning the tight focus is made possible because, as result of linearity, the difference matrix: $T_{1-2} = (T_1 - T_2)$ describes an AOTM of a virtual ultrasound focus that is obtained by subtracting the first ultrasound focus pressure field from the other (Fig3e). This difference acoustic pressure field is zero at a specific distance between the two ultrasound foci centers (Fig.3e). Dividing the sum of the two ultrasound foci (Fig.3d) by the difference ultrasound pressure field (Fig.3e) results in a sharp peak along at this distance (Fig.3f). The position of the sharp peak is controlled by the parameter $\alpha$. In practice, to take into account measurement noise the matrix inversion in the calculation of $A_\alpha$ is performed via a pseudo-inverse (+) with a proper regularization parameter[38].

Figure 3i-k present numerical results obtained by this approach, with a comparison to TRUE focusing (Fig.3g-h). Figure 4 presents results of a proof-of-principle experiment. It can be noticed that the focus obtained by decomposition of $A_\alpha$ (Fig.3i-k, Fig.4e-h) is not only smaller than the ultrasound focus (Fig.3(b-c,g-h), Fig.4a-b), but that it is also sharper than the foci obtained using SVD of a single AOTM (Fig.4d). Focusing using the joint information in the two matrices yields superior results to the ones obtained by considering each matrix separately since the peak of the virtual ultrasound focus obtained by dividing by $T_{1-\alpha 2}$ (Fig.3f) is sharper than the peak of each of the Gaussian foci (Figs.b-c). This result can be extended to allow scanning in two or three dimensions using a larger number of ultrasound foci, as was demonstrated by Judkewitz et al. in their TROVE work[25]. Interestingly, the variance maximization approach that is employed in TROVE via diagonalization of a spatial covariance matrix in the form of $T^H T$, can now be interpreted as an SVD of an AOTM, $T$, measured with random input modes. The AOTM thus also encompasses TROVE, in addition to TRUE and iTRUE, with the major advantage of the AOTM being that any positioning of the SLM and camera can be exploited, and there is no requirement for a DOPC system[31].

In the experimental results of Fig.4, the FWHM transverse dimensions of the ultrasound foci (Fig.4a-b) and the transverse dimensions of sharpest focus obtained SVD of $A_\alpha$ (Fig.2f) are 180μm±10μm, and 50μm±5μm, respectively. The axial dimensions of these foci are 200μm±20μm, and 60μm±5μm, respectively. Thus, the approach provides here a resolution increase of more than ×3.3.

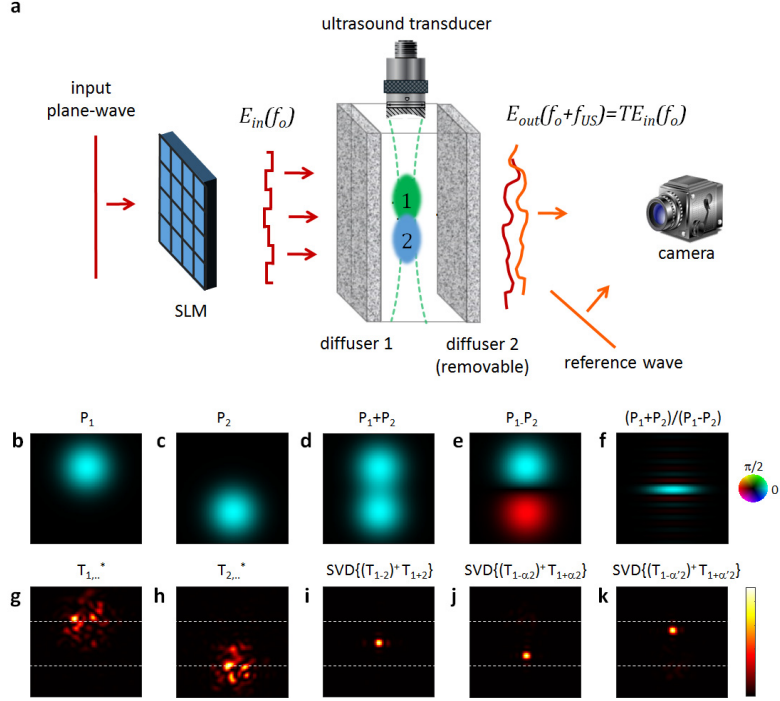

**Figure 3: Extending the AOTM to multiple ultrasound foci for optical focus scanning (numerical results): a,** measurement setup: two partly overlapping ultrasound foci (marked by '1' and '2' in green and blue, respectively) are used to measure two AOTMs, $T_1$ and $T_2$; **b,c,** ultrasound pressure amplitude distributions for the first ($P_1$) and second ($P_2$) foci; **d,e,** sum and differences of the ultrasound pressure fields; **f,** sum of the pressure fields divided by the difference between the pressure fields, displaying a sharp peak between the two ultrasound foci centers; **(g-k)** optical intensity distribution at the ultrasound focal plane for TRUE-focusing using $P_1$ or $P_2$ **(g-h)**, and focusing via SVD of a matrix $A^\alpha$ (Equation 4) formed by the weighted sum of $T_1+\alpha T_2$ divided by their weighted difference $T_1-\alpha T_2$, for different values of α: α=1 **(i)**, α=0.05 **(j)**, and α=20 **(k)**.

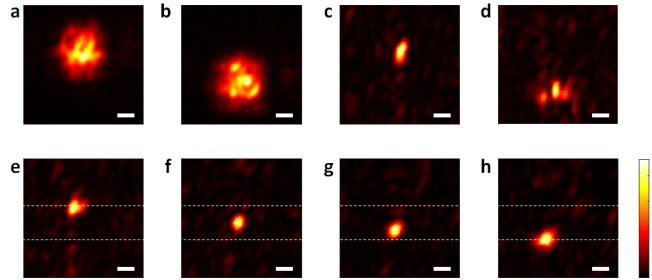

**Figure 4: Experimental focal scan by joint decomposition of AOTMs of two ultrasound foci: (a,b)** average intensity of the acoustically tagged light by each of the two ultrasound foci. **(c,d)** Focusing via SVD of $T_1$ **(c)** or $T_2$ **(d)** independently, as in Figs 1-2. **(e-h)** optical foci obtained by injecting the first singular vectors of the matrix $A^\alpha$ (Equation 4), with values of α of 0.05 **(e)**, 0.1 **(f)**, 1 **(g)**, and 10 **(h)**, demonstrating the ability to scan a tight optical focus between the two ultrasound foci. scale-bars, 90μm.

## Discussion

The AOTM provides both a tool for non-invasive high-resolution optical investigation of complex media, and a mathematical framework for analyzing any experiment involving linear acousto-optic interaction in complex media. Being a general framework, the state-of-the-art acousto-optics techniques could be revisited using the AOTM: any TRUE-focusing experiment is described by a single row of the AOTM, and the first singular vector of the AOTM gives the wavefront obtained after infinite iterations of iTRUE, and the TROVE-focusing wavefront using a single ultrasound focus.

An important advantage of the AOTM compared to TRUE, iTRUE and TROVE techniques, is that it does not require a DOPC system or any careful matching of input to output pixels, since the position of the SLM and camera is arbitrary. Another potential advantage is the possibility to have an asymmetry in the pixel numbers: measuring on a much larger number of output camera pixels than SLM pixels, as permitted by state-of-the-art cameras, with no penalty on the measurement time. However, compared to TRUE and iTRUE, the AOTM is considerably slower, since the measurements of the controlled input modes are sequential.

We have shown that the singular values of the AOTM provide information about the shape and size of the ultrasound focus (Fig.1c, Supplementary Figure S3), and that the AOTM singular vectors concentrate light at different locations around the ultrasound focus center. Interestingly, the operator $T^H T$ that is diagonalized to perform the SVD of the AOTM has been studied extensively in the context of time-reversal ultrasound experiments in reflection, and was termed the time-reversal operator (TRO)[39-41]. The time-reversed acoustics analogue of the spatially extended acousto-optic tagging focus considered here, is an extended reflecting target[42, 43]. It was proved that the number of significant eigenvalues of the TRO corresponds to the number of modes (or resolution cells) contained in the target area, and that, in the Fraunhofer approximation, these modes are given by prolate spheroidal wave-functions[43], functions that have been studied in e.g. information theory as a basis of band-limited and time-limited signals.

As a tool, we have shown how the AOTM can be measured with a single or multiple ultrasonic beams, and how it can be used to provide sub acoustic-resolution focusing. Scanning a sharp focus inside the sample is valuable for imaging applications, and for studying various models of optical propagation in diffusive media in three-dimensions, such as the intensity distribution of transmission eigenchannels[36], Anderson localized modes, and the memory-effect inside (rather than through) volumetric samples[44, 45]. The extension to two acoustic beams (Figs.3-4) can be scaled up to multiple acoustic regions, allowing a large field of view. However, when large volumes are considered, the computational resources for storing the high-dimensional (3D) AOTM are expected to be very high (and the acquisition process time consuming). This is because the AOTM dimension with only one ultrasound focus is the multiplication of the number of input optical modes, $N_{SLM}$, on a two-dimensional grid by the number of optical modes, $M$, on an output two-dimensional grid: $M{\times}N_{SLM}$. An AOTM covering a large 3D volume will require $L$ focused ultrasound pulses, where $L$ reflects a three-dimensional (x,y,z) scan, resulting in a matrix with $LxMxN$ elements. The focused ultrasound beams may be replaced by ultrasonic plane waves, which can be digitally combined coherently to synthetize various ultrasonic foci in 3D[46, 47]. The AOTM could also be extended to non-monochromatic optical illumination, providing spectral and temporal information on the medium's response, This could be realized by using ultrashort optical pulses[48], or spectrally tunable sources[49], but would results in yet higher dimensionality of the AOTM.

## Methods

**Experiments.** The full experimental setup is described in detail in Supplementary Section 1: A long-coherence continuous-wave (cw) laser at a center wavelength of 810nm was used as the light source. The laser is a single longitudinal mode, tunable, extended cavity, semiconductor laser, producing up to 1.5Watt. The maximum power used in our experiments was 300mW. The laser was provided by DTU Fotonik, Denmark (see acknowledgements). To measure the ultrasonically-modulated light, the laser beam is split to two arms of an interferometer. At the first arm, the laser beam illuminates an SLM (Holoeye Pluto), which is imaged on the first diffuser. A spherically-focused ultrasonic transducer (V319-SU-F0.75-IN-PTF, Olympus; 15MHz central frequency, 0.75" focal length, 0.5" element diameter), emits 133ns long pulses at a center frequency of 15MHz inside a water filled glass cuvette. At the second, reference, arm, two acousto-optic modulators (MT-80, AA-Optoelectronics) are used to frequency shift and time gate the reference arm signal, to produce a 133ns pulse with a central frequency that is shifted by 15MHz+(5KHz/4) from the original laser frequency. The pulsed reference beam is focused on a mirror placed next to the output plane of the medium (the second diffuser) and is combined with the first beam on a fast camera (Photron Fastcam SA4) operating at $f_{cam}=5,000$ frames per second. To measure the weak ultrasonically-tagged field in the presence of the intense untagged background, a double-heterodyne holographic technique was employed[32], combining off-axis and phase-shifting interferometry. To maximize the measurements' signal to noise the SLM was imaged on the first diffuser surface, and a Hadamard input basis was used to measure the AOTM[6]. To compensate for any slow phase drifts of the reference arm, a flat-phase mask was displayed on the SLM before each input mode was injected, and the phase of the output field measured with a flat phase input was subtracted from the phase of each measured output field. The total acquisition time for a single AOTM with 3072 input modes (6144 displayed phase patterns) was limited by the SLM refresh rate (~6Hz), and was 18 minutes. An iris placed in front of the first diffuser was used to control the speckle size at the target plane. Triggering of all instruments is detailed in Supplementary Figure 2. To minimize the fraction of the unshaped light from the SLM limited fill-factor, a ruled grating phase pattern was displayed on the SLM and only the first diffraction order was collected and injected into the sample. The displayed phase-pattern on the SLM was composed of macro pixels of 20×20 pixels, displaying the phase sum of the ruled grating and the desired input phase. For direct inspection of the optical fields at the ultrasound focal plane, the second diffuser was removed, and an imaging lens was placed between the camera and the scattering medium.

**Data Analysis.** The experimentally measured AOTMs had 3072 input modes (Hadamard basis dimension). For each input mode a camera image of the output field with a resolution of 512×352 pixels (Fig.2) or 320×240 (Fig.4) was acquired. To minimize the computational memory requirements, considering the speckle size on the camera, one out of three camera pixel in each dimension was taken as an output mode (×3 under-sampling). For the results of Figs.1-2, SVD was performed in Matlab (Mathworks). For the results of Figs.3-4 SVD was performed on the matrix $A_\alpha = (T_{1-\alpha2}^H T_{1-\alpha2})^+ (T_{1+\alpha2}^H T_{1+\alpha2})$, where $^+$ is a Tikhonov regularized pseudo inverse[38]. Since the number of columns and rows of $T$ can be substantially different, the SVD of $B_\alpha = (T_{1-\alpha2} T_{1-\alpha2}^H)^+ (T_{1+\alpha2} T_{1+\alpha2}^H)$ can be used as well to calculate the optimal focusing input vectors, reducing the size of the analyzed matrix[25].

**Simulations**. Two random matrices with Gaussian independent and identically distributed (i.i.d) random amplitude entries and random i.i.d phase distribution from zero to $2\pi$ were generated to describe the all-optical transmission matrix between: (1) the SLM plane and the ultrasound focal plane, $T_{SLM-US}$, and (2) the ultrasound focal plane to the camera, $T_{US-CAM}$. The propagation from the SLM to the camera was simulated by multiplying $T_{SLM-US}$ by the input-field $E_{in}$. The ultrasound tagging was simulated by multiplying the field $T_{US-CAM}E_{in}$ pixel-by-pixel by a Gaussian-shaped ultrasound pressure amplitude $P_{US}$. The results was propagated to the camera by multiplying with $T_{US-CAM}$. The number of input and output modes were 4096 and 2048, respectively.

**References**

1. Yaqoob, Z., Psaltis, D., Feld, M.S. & Yang, C. Optical phase conjugation for turbidity suppression in biological samples. *Nat Photon* **2**, 110-115 (2008).
2. Fink, M. Time reversal of ultrasonic fields. I. Basic principles. *IEEE transactions on ultrasonics, ferroelectrics, and frequency control* **39**, 555-566 (1992).
3. Mosk, A.P., Lagendijk, A., Lerosey, G. & Fink, M. Controlling waves in space and time for imaging and focusing in complex media. *Nat Photon* **6**, 283-292 (2012).
4. Horstmeyer, R., Ruan, H. & Yang, C. Guidestar-assisted wavefront-shaping methods for focusing light into biological tissue. *Nature photonics* **9**, 563-571 (2015).
5. Vellekoop, I.M. & Mosk, A.P. Focusing coherent light through opaque strongly scattering media. *Opt Lett* **32**, 2309-2311 (2007).
6. Popoff, S.M. et al. Measuring the transmission matrix in optics: an approach to the study and control of light propagation in disordered media. *Phys Rev Lett* **104**, 100601 (2010).
7. Vellekoop, I.M., Lagendijk, A. & Mosk, A.P. Exploiting disorder for perfect focusing. *Nat Photon* **4**, 320-322 (2010).
8. Vellekoop, I.M. & Aegerter, C.M. Scattered light fluorescence microscopy: imaging through turbid layers. *Opt Lett* **35**, 1245-1247 (2010).
9. Katz, O., Small, E. & Silberberg, Y. Looking around corners and through thin turbid layers in real time with scattered incoherent light. *Nature Photonics* **6**, 549-553 (2012).
10. Aulbach, J., Gjonaj, B., Johnson, P.M., Mosk, A.P. & Lagendijk, A. Control of Light Transmission through Opaque Scattering Media in Space and Time. *Physical Review Letters* **106**, 103901 (2011).
11. Katz, O., Small, E., Bromberg, Y. & Silberberg, Y. Focusing and compression of ultrashort pulses through scattering media. *Nat Photon* **5**, 372-377 (2011).
12. Small, E., Katz, O., Guan, Y. & Silberberg, Y. Spectral control of broadband light through random media by wavefront shaping. *Opt. Lett.* **37**, 3429-3431 (2012).
13. Guan, Y., Katz, O., Small, E., Zhou, J. & Silberberg, Y. Polarization control of multiply scattered light through random media by wavefront shaping. *Opt. Lett.* **37**, 4663-4665 (2012).
14. Popoff, S., Goetschy, A., Liew, S., Stone, A.D. & Cao, H. Coherent control of total transmission of light through disordered media. *Physical review letters* **112**, 133903 (2014).

15. Rotter, S. & Gigan, S. Light fields in complex media: Mesoscopic scattering meets wave control. *Reviews of Modern Physics* **89**, 015005 (2017).
16. Vellekoop, I.M., van Putten, E.G., Lagendijk, A. & Mosk, A.P. Demixing light paths inside disordered metamaterials. *Opt Express* **16**, 67-80 (2008).
17. Hsieh, C.-L., Pu, Y., Grange, R. & Psaltis, D. Digital phase conjugation of second harmonic radiation emitted by nanoparticles in turbid media. *Opt. Express* **18**, 12283-12290 (2010).
18. Tang, J., Germain, R.N. & Cui, M. Superpenetration optical microscopy by iterative multiphoton adaptive compensation technique. *Proceedings of the National Academy of Sciences* (2012).
19. Katz, O., Small, E., Guan, Y. & Silberberg, Y. Noninvasive nonlinear focusing and imaging through strongly scattering turbid layers. *Optica* **1**, 170-174 (2014).
20. Kong, F. et al. Photoacoustic-guided convergence of light through optically diffusive media. *Opt. Lett.* **36**, 2053-2055 (2011).
21. Chaigne, T. et al. Controlling light in scattering media non-invasively using the photoacoustic transmission matrix. *Nat Photon* **8**, 58-64 (2014).
22. Conkey, D.B. et al. Super-resolution photoacoustic imaging through a scattering wall. *Nature communications* **6** (2015).
23. Xu, X., Liu, H. & Wang, L.V. Time-reversed ultrasonically encoded optical focusing into scattering media. *Nat Photon* **5**, 154-157 (2011).
24. Wang, Y.M., Judkewitz, B., DiMarzio, C.A. & Yang, C. Deep-tissue focal fluorescence imaging with digitally time-reversed ultrasound-encoded light. *Nat Commun* **3**, 928 (2012).
25. Judkewitz, B., Wang, Y.M., Horstmeyer, R., Mathy, A. & Yang, C. Speckle-scale focusing in the diffusive regime with time reversal of variance-encoded light (TROVE). *Nat Photon* **7**, 300-305 (2013).
26. Si, K., Fiolka, R. & Cui, M. Fluorescence imaging beyond the ballistic regime by ultrasound-pulse-guided digital phase conjugation. *Nat Photon* **6**, 657-661 (2012).
27. Si, K., Fiolka, R. & Cui, M. Breaking the spatial resolution barrier via iterative sound-light interaction in deep tissue microscopy. *Sci. Rep.* **2** (2012).
28. Ruan, H., Jang, M., Judkewitz, B. & Yang, C. Iterative Time-Reversed Ultrasonically Encoded Light Focusing in Backscattering Mode. *Sci. Rep.* **4** (2014).
29. Tay, J.W., Lai, P., Suzuki, Y. & Wang, L.V. Ultrasonically encoded wavefront shaping for focusing into random media. *Scientific Reports* **4**, 3918 (2014).
30. Elson, D.S., Li, R., Dunsby, C., Eckersley, R. & Tang, M.-X. Ultrasound-mediated optical tomography: a review of current methods. *Interface focus* **1**, 632-648 (2011).
31. Cui, M. & Yang, C. Implementation of a digital optical phase conjugation system and its application to study the robustness of turbidity suppression by phase conjugation. *Opt. Express* **18**, 3444-3455 (2010).
32. Atlan, M., Forget, B.C., Ramaz, F., Boccara, A.C. & Gross, M. Pulsed acousto-optic imaging in dynamic scattering media with heterodyne parallel speckle detection. *Optics Letters* **30**, 1360-1362 (2005).
33. Vellekoop, I.M. & Mosk, A.P. Universal optimal transmission of light through disordered materials. *Phys Rev Lett* **101**, 120601 (2008).
34. Kim, M. et al. Maximal energy transport through disordered media with the implementation of transmission eigenchannels. *Nat Photon* **6**, 581-585 (2012).
35. Sarma, R., Yamilov, A.G., Petrenko, S., Bromberg, Y. & Cao, H. Control of energy density inside a disordered medium by coupling to open or closed channels. *Physical review letters* **117**, 086803 (2016).
36. Liew, S.F., Popoff, S.M., Mosk, A.P., Vos, W.L. & Cao, H. Transmission channels for light in absorbing random media: from diffusive to ballistic-like transport. *Physical Review B* **89**, 224202 (2014).
37. Popoff, S.M. et al. Exploiting the Time-Reversal Operator for Adaptive Optics, Selective Focusing, and Scattering Pattern Analysis. *Physical Review Letters* **107**, 263901 (2011).
38. Popoff, S., Lerosey, G., Fink, M., Boccara, A.C. & Gigan, S. Image transmission through an opaque material. *Nat Commun* **1**, doi:10 1038/ncomms1078 (2010).
39. Prada, C. & Fink, M. Eigenmodes of the time reversal operator: A solution to selective focusing in multiple-target media. *Wave Motion* **20**, 151-163 (1994).
40. Prada, C., Manneville, S., Spoliansky, D. & Fink, M. Decomposition of the time reversal operator: Detection and selective focusing on two scatterers. *The Journal of the Acoustical Society of America* **99**, 2067-2076 (1996).

41. Prada, C. & Thomas, J.-L. Experimental subwavelength localization of scatterers by decomposition of the time reversal operator interpreted as a covariance matrix. *The Journal of the Acoustical Society of America* **114**, 235-243 (2003).
42. Komilikis, S., Prada, C. & Fink, M. in Ultrasonics Symposium, 1996. Proceedings., 1996 IEEE, Vol. 2 1401-1404 (IEEE, 1996).
43. Robert, J.-L. & Fink, M. The prolate spheroidal wave functions as invariants of the time reversal operator for an extended scatterer in the Fraunhofer approximation. *The Journal of the Acoustical Society of America* **125**, 218-226 (2009).
44. Freund, I., Rosenbluh, M. & Feng, S. Memory Effects in Propagation of Optical Waves through Disordered Media. *Physical Review Letters* **61**, 2328-2331 (1988).
45. Judkewitz, B., Horstmeyer, R., Vellekoop, I.M., Papadopoulos, I.N. & Yang, C. Translation correlations in anisotropically scattering media. *Nat Phys* **11**, 684-689 (2015).
46. Laudereau, J.-B., Grabar, A.A., Tanter, M., Gennisson, J.-L. & Ramaz, F. Ultrafast acousto-optic imaging with ultrasonic plane waves. *Optics Express* **24**, 3774-3789 (2016).
47. Montaldo, G., Tanter, M., Bercoff, J., Benech, N. & Fink, M. Coherent plane-wave compounding for very high frame rate ultrasonography and transient elastography. *IEEE transactions on ultrasonics, ferroelectrics, and frequency control* **56**, 489-506 (2009).
48. Choi, Y. et al. Measurement of the time-resolved reflection matrix for enhancing light energy delivery into a scattering medium. *Physical review letters* **111**, 243901 (2013).
49. Andreoli, D. et al. Deterministic control of broadband light through a multiply scattering medium via the multispectral transmission matrix. *Scientific reports* **5** (2015).

**Acknowledgements**

The authors thank Mingjun Chu and Paul Michael Petersen, Technical University of Denmark, for providing the long-coherence laser source, and Benjamin Judkewitz for discussions.

This project has received funding from the European Research Council (ERC) under the European Union's Horizon 2020 research and innovation program (grants no. 278025, 677909), and LABEX WIFI (Laboratory of Excellence within the French Program "Investments for the Future") under references ANR-10-LABX-24. O.K. was supported by a Marie Curie intra-European fellowship (IEF) and Azrieli Faculty Fellowship. S.G. acknowledge support from the Institut Universitaire de France (IUF).
**Author contributions**

M.F. conceived the idea. O.K. developed the idea, performed modeling and numerical simulations, built the set-up, collected data, and performed data analysis. All authors contributed to the design of the experiments and to the writing of the manuscript.

# Controlling light in complex media beyond the acoustic diffraction-limit using the acousto-optic transmission matrix - Supplementary Information

## 1. Experimental setup

A sketch of the optical experimental setup is given in Supplementary Figure 1. The scheme for the triggering and electronic connections is given in Supplementary Figure 2.

The optical setup is based on double-heterodyne detection scheme for acousto-optic tomography[1]. This detection scheme combines off-axis holography with phase shifting holography to provide sensitive detection of the weak ultrasonically modulated signal over the strong unmodulated background[2].

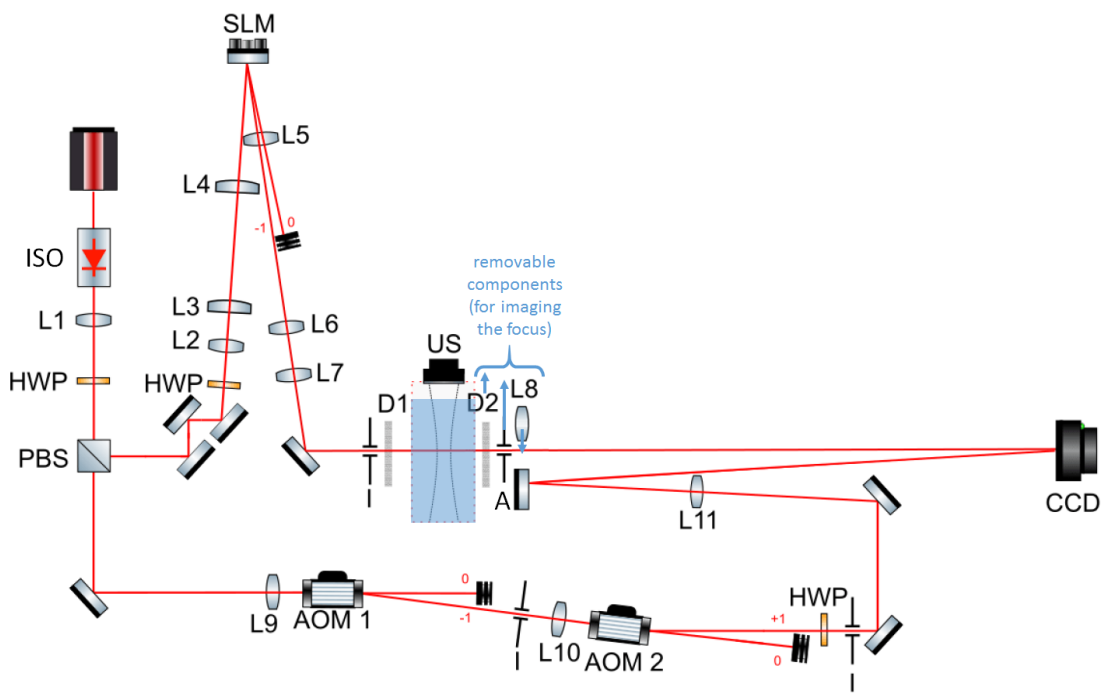

Supplementary Figure S1, Experimental set-up. (ISO – Isolator, L# - lens, HWP – half wave plate, PBS – polarization beam splitter, AOM – acousto-optic modulator, I – iris, D# - diffuser, A – rectangular aperture, CCD – camera, US – ultrasound transducer, SLM – spatial light modulator)

The light source is a long-coherence continuous-wave infrared semiconductor laser. It is a compact (8×4×6cm) tunable extended cavity single longitudinal mode laser, centered at 810nm. The laser was provided by DTU Fotonik, Denmark (see acknowledgements). The compact laser can provide up to 1.5 Watt, without the need of an optical amplifier. 300mW average power was used in our experiments. The laser beam passes an isolator, is collimated by L1 and split to two arms of an interferometer by a polarization beam splitter. The relative powers in the two arms are controlled by a half wave plate. The beam paths in the signal and reference arms are as follows:

The beam at the signal arm is magnified using a telescope made up of lens L2 and two cylindrical lenses L3 and L4 (at different axes) to match the SLM dimensions. The zero diffraction order of the light shaped by the SLM is cut, and a telescope made out of lenses L6 and L7 images the SLM on the first diffuser, D1, using the light from the first diffraction order. An iris, I, placed before the first diffuser controls the speckle size at the target plane. The light passes a water-filled cuvette, where the ultrasound beam is focused at. A second diffuser (D2) is placed at the output facet of the cuvette. The scattered light passes an aperture (A) that matches the speckle size on the camera to the camera resolution.

The reference arm passes two acousto-optic modulators, and is polarization rotated by a HWP to match the scattered light polarization at the signal arm. The reference arm is focused by L11 at the plane of the aperture placed after the exit of the second diffuser D2, and is reflected by a mirror to be combined at a small angle on the camera. To directly image the focal plane, the second diffuser D2 and the aperture, A, are removed and an imaging lens L8 is used to image the focal plane on the camera.

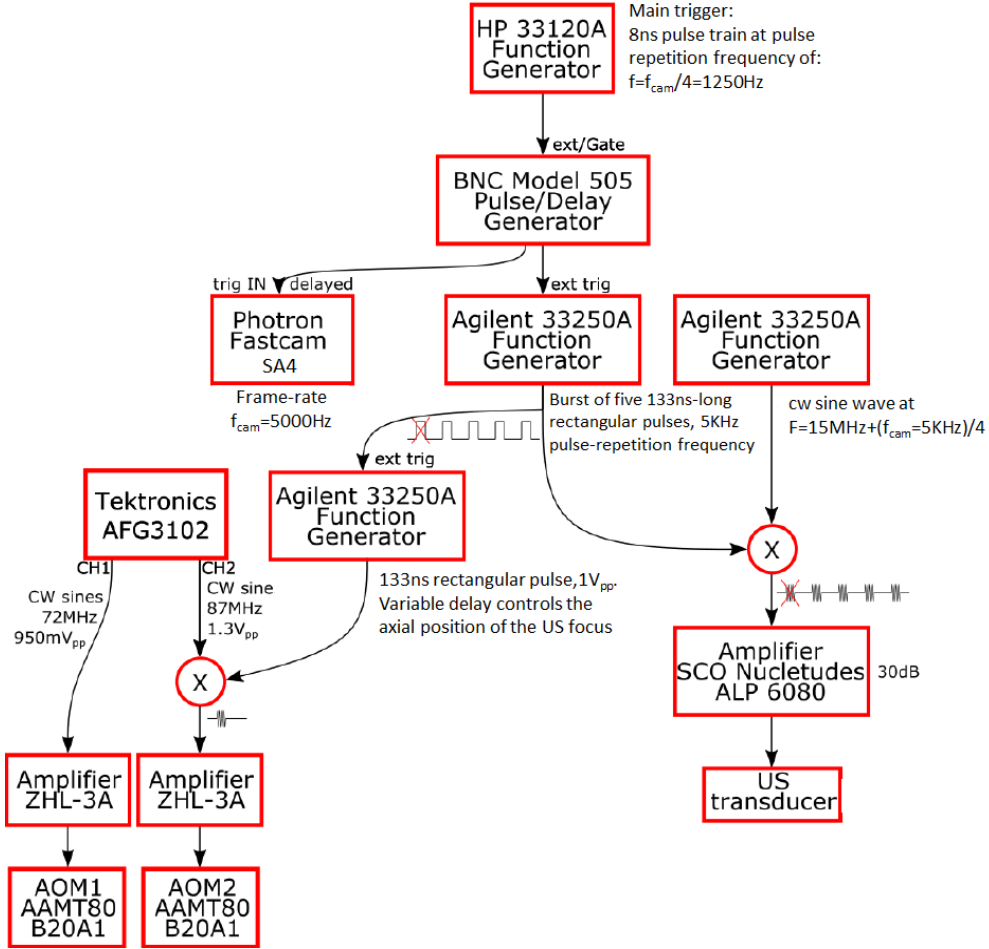

Supplementary Figure S2: Electronic connections and triggering scheme

## 2. TRUE focusing as phase conjugation of an AOTM row

The AOTM, $T$, describes the propagation of light from the SLM plane, through the ultrasound focus, and to the camera plane. As result of optical reciprocity, the matrix describing the propagation of light from the m-th camera pixel to the SLM plane is $T^T$. Thus, if light is injected into the medium from the $m^{th}$ mode *on the camera side*, it will result in an output field *at the SLM plane* given by: $T^T E_{in,cam}$, i.e. by the $m^{th}$ column of $T^T$. If this field is used for TRUE focusing, TRUE focusing will be obtained by displaying on the SLM the phase conjugate of this field, i.e. the phase conjugate of the $m^{th}$ column of $T^T$, which is the m-th row of $T$, given by $(t_{m,*})^*$. Thus, displaying the m-th row of the AOTM, $T$, is equivalent to performing TRUE-focusing for the light input from the m-th camera pixel.

## 3. Equivalence between the first singular vector of the AOTM and infinite iterations of iterative time-reversed ultrasound encoded (iTRUE) focusing

Here we analyze iterative TRUE (iTRUE) iterations under the AOTM framework. Following the definition of the AOTM as the relation between the input field to the medium and measured ultrasonically tagged field (Eq.1), the measured field on the camera in the first iTRUE iteration is given by:

$$E^{out}(f_o + f_{US}) = TE^{in}(f_o) \quad (S1)$$

In the first iterations this field is phase-conjugated and injected back into the medium at the original laser frequency[3]. Due to optical reciprocity, the propagation of this phase-conjugated optical field back through the medium and through the ultrasonic focus is given by $T^T$. Thus, the optical field measured at the output in the first iTRUE iteration, which is the input field to the second iTRUE iteration is given by:

$$E^{out,1}(f_o + f_{US}) = T^T \left(TE^{in}(f_o)\right)^* = T^T T^* \left(E^{in}(f_o)\right)^* \quad (S2)$$

At the second iTRUE iteration, $E^{out,1}$ is phase-conjugated to provide:

$$E^{in,2} = \left(E^{out,1}\right)^* = \left(T^T T^* \left(E^{in}(f_o)\right)^*\right)^* = (T^H T) E^{in}(f_o) \quad (S3)$$

Where $T^H$ is the Hermitian conjugate of the matrix $T$.

Thus, the field conjugated at the $(2k)^{th}$ iTRUE iteration is given by:

$$E^{in,2k} = (T^H T)^k E^{in} \quad (S4)$$

In ultrasound $T^H T$ was termed the 'time reversal operator' (TRO) [4,5]. Theoretically, the tightest optical focus will be obtained after performing an infinite number of iterations, k→∞. Following Supp.Eq.4, the injected field in this k→∞ iteration would be given by the matrix $T^H T$ (the TRO) taken to the $k^{th}$→∞ power. Thus, iTRUE is expected to converge to the eigenvector of $T^H T$ having the largest eigenvalue, i.e. the first singular vector of $T$.

The result of the iTRUE power-iterations, is analogous to the formation of the lowest-loss lasing mode in a laser cavity. Specifically, the result is obtained under the assumption that the input field is decomposed into all eigenvectors, $V_i$, of the matrix $T^H T$. Ordering these modes by the amplitude of their eigenvalues $\lambda_i$, such that $(T^H T)V_i = \lambda_i V_i$ with $\lambda_i \geq \lambda_{i-1}$, and writing the input field as $E^{in} = \sum_i a_i V_i$ the injected field in the k→∞ iteration then becomes:

$$\lim_{k\to\infty}\left((T^H T)^k E^{in}\right) = \lim_{k\to\infty}((T^H T)^k \sum_i a_i V_i) = \lim_{k\to\infty}\left(\sum_i \lambda_i^k V_i\right) \propto V_1 \quad (S5)$$

Thus, for optimal optical focusing one needs to send into the medium the eigenvector $V_1$, of the time-reversal operator $T^H T$, with the largest eigenvalue, $\lambda_1$.

The process of computing the first eigenvector of $T^H T$ is equivalent to computing the first *singular* vector of the AOTM, $T$. Thus, all that is required to find the optimal focusing wavefront, is to perform a singular value decomposition (SVD) of the AOTM. SVD of a matrix $T$ is given by $T = USV^*$, where $S$ is a rectangular diagonal matrix containing the real positive singular values, $\mu_i$, in descending order, and $U$ and $V$ are unitary matrices whose columns corresponds to the output and input singular vectors, $U_i$ and $V_i$, respectively. Each input singular vector $V_i$ corresponds to the input field (at the SLM plane) that corresponds to the $i^{th}$ singular value, $\lambda_i$. The corresponding output singular vector $U_i$ is expected resulting field at the camera plane.

## 4. Singular value decomposition of Gaussian vs. a top-hat tagging function

Supplementary Figure S3 presents the singular values obtained via SVD of two simulated AOTMs. The first (Supplementary Fig.S3a) is an AOTM measured with a two-dimensional Gaussian shaped ultrasonic focus, close to the one expected to be achieved in practice, and the same one used to simulate the results of Figure 1 in the main text. The second (Supplementary Fig.S3a) is for a theoretical tagging ultrasound focus having a 'top-hat' circular shape, with sharp defined edges. It can be observed the singular values sharply fall off for the top-hat shaped-target, as expected[6], where the singular values for the Gaussian shaped focus fall gradually. Supplementary Figures 4-5 display the results of focusing using the singular vectors corresponding to the first 162 singular values presented in Supplementary Figure 3, for both the Gaussian tagging beam (Supplementary Figure S4, Figure 1), and the top-hat beam (Supplementary Figure 5)

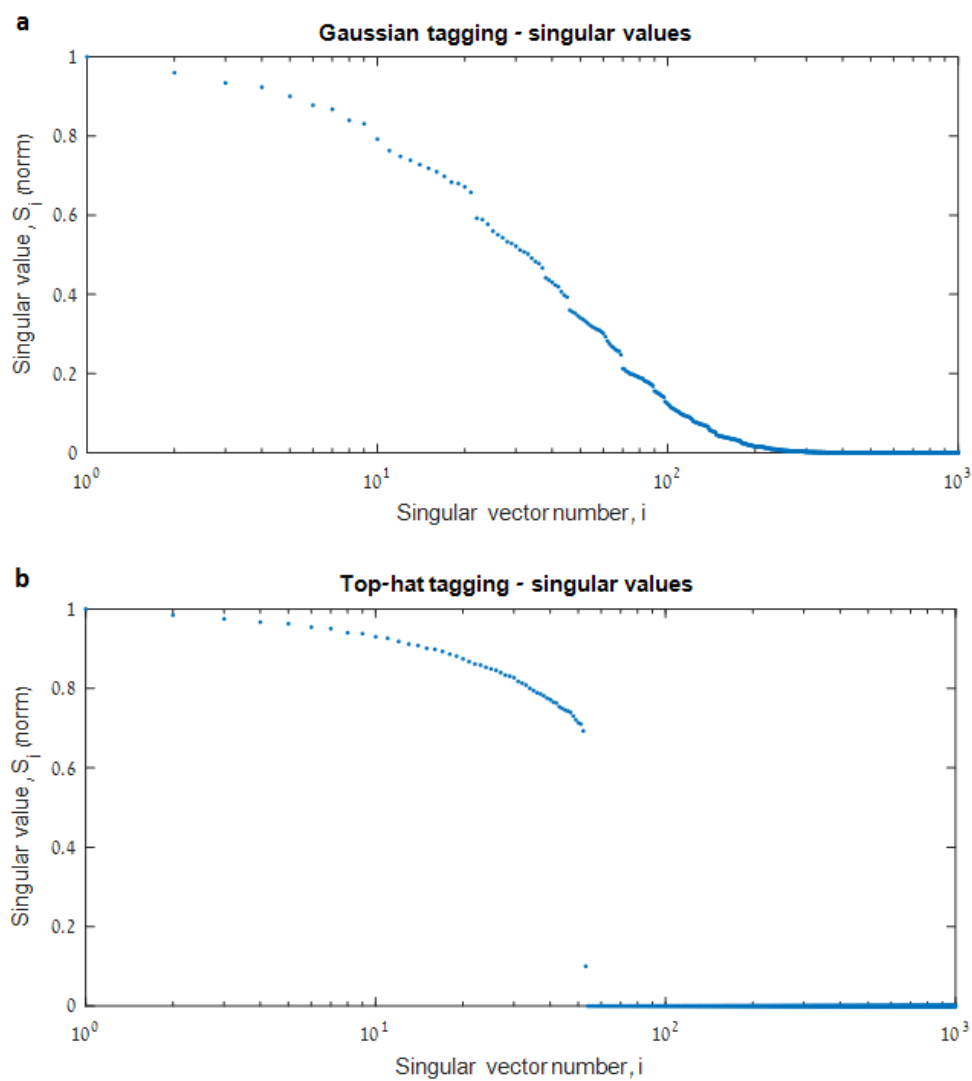

Supplementary Figure S3: Singular values of a Gaussian ultrasound focus (a) and a top-hat circular shaped tagging area (b)

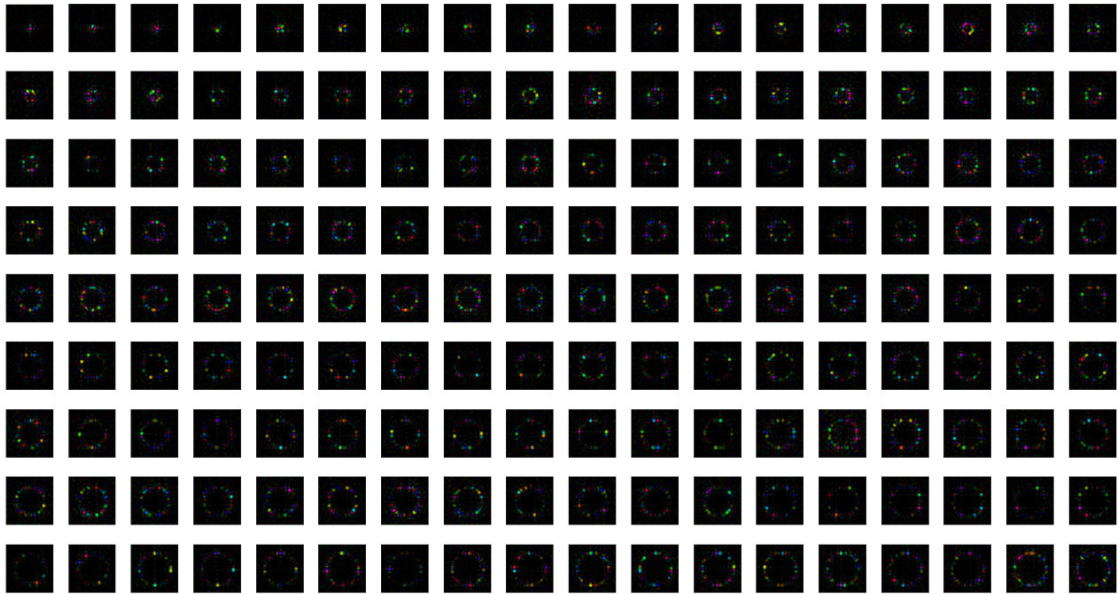

Supplementary Figure S4: Optical fields obtained at the acoustic focal plane when focusing with first 162 singular vectors of a Gaussian tagging area

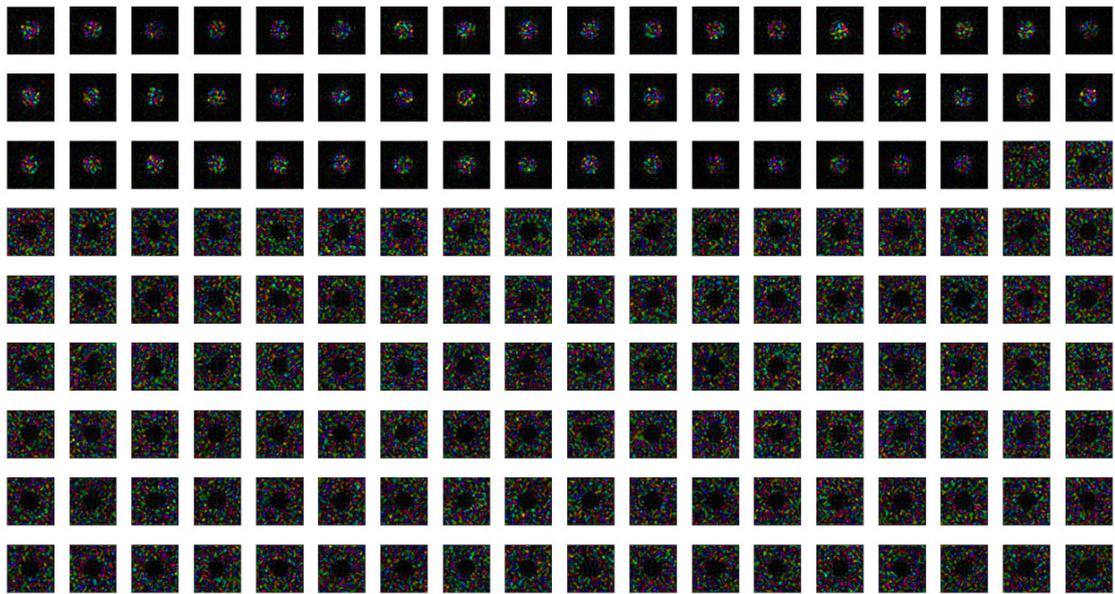

Supplementary Figure S5: Optical fields obtained at the acoustic focal plane when focusing with first 162 singular vectors of a 'top-hat' circular tagging area

# 5. Estimating the number of degrees of control (SLM pixels) required for single speckle scale focusing

Here we give a rough estimation for a lower bound on the number of controlled SLM pixels ($N_{SLM}$), i.e. the number of AOTM columns, that is required for focusing to a single speckle grain using the SVD of the AOTM. The derivation assumes 2D Gaussian-shape tagging.

To be able to focus on a single speckle of width $\sigma_{speckle}$ located at the center of the ultrasound focus via SVD of the AOTM, the energy transmitted through the complex sample using this first singular vector, $V_1$, needs to be higher than the energy of the second singular vector, which originates from focusing to an adjacent speckle, $V_2$.

Assuming an ultrasound focus pressure distribution having a Gaussian profile: $P_{US}(r) \propto \exp\left\{-\left(\frac{r}{\sigma_{US}}\right)^2\right\}$, this task becomes more difficult the smaller is the speckle grain size compared to the ultrasound focus size $\sigma_{US}$.

Given that for a phase only wavefront shaping, the intensity enhancement of a single speckle grain is given by[7]: $\eta = \frac{\pi}{4} N_{SLM}$, the energy transmitted through the ultrasound focus when the first singular vector is injected into the medium should be:

$$E_1 \propto \eta E_{speckle} \cdot P_{US}(r=0) + B_1 \approx \eta E_{speckle} + B_1 \quad (S6)$$

Where $E_{speckle}$ is the average energy of a single speckle grain, $\eta E_{speckle}$ is the energy of the wavefront-shaped speckle grain at the focus, and $B_1$ is a background term originating from the $N_{speckles} \approx (\sigma_{US}/\sigma_{speckle})^2$ speckle grains that are contained in the ultrasound focus.

Using the second singular vector, $V_2$, the total energy transmitted through the ultrasound focus can be approximated by considering a wavefront shaped intensity-enhanced speckle that is located $\Delta r \approx \sigma_{speckle}/2$ off the center of the ultrasound focus:

$$E_2 \propto \eta E_{speckle} \cdot P_{US}(r = \frac{\sigma_{speckle}}{2}) + B_2 \approx \eta E_{speckle}\left(1 - \left(\frac{\sigma_{speckle}}{2\sigma_{US}}\right)^2\right) + B_2 \quad (S7)$$

Where the ultrasound pressure distribution was approximated assuming the speckle grain size is considerably smaller than the ultrasound focus diameter.

To be able to focus on the single speckle at the center of the ultrasound focus, it is necessary that the first singular value (proportional to $E_1$) should be larger than the second one, which is proportional to $E_2$:

$$E_1 > E_2 \quad (S8)$$

$$\eta E_{speckle} + B_1 > E_{speckle}\left(1 - \left(\frac{\sigma_{speckle}}{2\sigma_{US}}\right)^2\right) + B_2 \quad (S9)$$

$$\eta E_{speckle}\left(\frac{\sigma_{speckle}}{2\sigma_{US}}\right)^2 > B_2 - B_1 \quad (S10)$$

Each of the background terms $B_1, B_2$ in equation (S10) is the sum of the $N_{speckles}$ contained within the ultrasound focus in the given speckle realization, and can be approximated by:

$$B_i \approx E_{speckle}\left(N_{speckles} \pm \sqrt{N_{speckles}}\right) \approx E_{speckle}(\sigma_{US}/\sigma_{speckle})^2 \pm E_{speckle}(\sigma_{US}/\sigma_{speckle}) \quad (S11)$$

Plugging these background terms difference into equation (S10) yields the condition:

$$\eta E_{speckle}\left(\frac{\sigma_{speckle}}{2\sigma_{US}}\right)^2 >\sim \sqrt{2} E_{speckle}\left(\frac{\sigma_{US}}{\sigma_{speckle}}\right) \quad (S12)$$

Plugging $\eta = \frac{\pi}{4} N_{SLM}$ gives the final condition:

$$N_{SLM} >\sim \frac{16}{\pi}\sqrt{2}\left(\frac{\sigma_{US}}{\sigma_{speckle}}\right)^3 \qquad (S13)$$

This result suggests that the number of required controlled input modes (number of columns in the AO-TM) is expected to *scale cubically* with the ratio between the ultrasound focus diameter and speckle diameter. For example, when $\left(\frac{\sigma_{US}}{\sigma_{speckle}}\right) = 5$, Equation (S13) gives $N_{SLM}>900$.

The above derivation assumed noise-free measurements. Measurement noise may be considered by adding an additional noise term to the right-hand side of equation S12, giving the final condition for the number of controlled modes:

$$N_{SLM} >\sim \frac{4}{\pi}\left(\sqrt{2}\left(\frac{\sigma_{US}}{\sigma_{speckle}}\right)^3 + \sigma_{noise}\left(\frac{\sigma_{US}}{\sigma_{speckle}}\right)^2\right) \qquad (S14)$$

## References

1. Atlan, M., Forget, B.C., Ramaz, F., Boccara, A.C. & Gross, M. Pulsed acousto-optic imaging in dynamic scattering media with heterodyne parallel speckle detection. *Optics Letters* **30**, 1360-1362 (2005).
2. Gross, M. & Atlan, M. Digital holography with ultimate sensitivity. *Optics letters* **32**, 909-911 (2007).
3. Ruan, H., Jang, M., Judkewitz, B. & Yang, C. Iterative Time-Reversed Ultrasonically Encoded Light Focusing in Backscattering Mode. *Sci. Rep.* **4** (2014).
4. Prada, C. & Fink, M. Eigenmodes of the time reversal operator: A solution to selective focusing in multiple-target media. *Wave Motion* **20**, 151-163 (1994).
5. Tanter, M., Thomas, J.L. & Fink, M. Time reversal and the inverse filter. *J Acoust Soc Am* **108**, 223-234 (2000).
6. Komilikis, S., Prada, C. & Fink, M. in Ultrasonics Symposium, 1996. Proceedings., 1996 IEEE, Vol. 2 1401-1404 (IEEE, 1996).
7. Vellekoop, I.M. & Mosk, A.P. Focusing coherent light through opaque strongly scattering media. *Opt Lett* **32**, 2309-2311 (2007).